\documentclass[preprint, 12pt, 3p, authoryear]{elsarticle}
\newdimen\Lmargin
\newdimen\Rmargin
\Rmargin=\paperwidth \advance\Rmargin by-\Lmargin \advance\Rmargin by-\textwidth
\ifdim\Lmargin>\Rmargin \Rmargin=\Lmargin \fi

\usepackage{amssymb}
\usepackage{graphicx}
\usepackage[colorlinks=true]{hyperref}
\usepackage{multirow}
\usepackage{booktabs}
\usepackage{natbib}
\usepackage{tabularx}
\usepackage{booktabs}
\usepackage{makecell}
\usepackage{amsmath}
\usepackage{lscape}
\usepackage{utfsym}
\usepackage{longtable}
\usepackage{caption}
\usepackage{amsthm}

\begin{document}

\begin{frontmatter}
\newtheorem{thm}{Theorem}
\newtheorem{deff}{Definition}
\newtheorem{pro}{Proposition}
\newtheorem{prot}{Property}
\newtheorem{lem}{Lemma}

\date{August 22, 2024}


\title{A Novel $\delta$-SBM-OPA Approach for Policy-Driven Analysis of Carbon Emission Efficiency under Uncertainty and its Application in Chinese Industry}

\author[inst1]{Shutian Cui}
\ead{C19061092@163.com}
\affiliation[inst1]{organization={School of Economics and Management, Northwest A$\&$F University},
            city={Yangling},
            postcode={712100}, 
            country={China}}
            
\author[inst2]{Renlong Wang}
\ead{13127073530@163.com}
\affiliation[inst2]{organization={School of Emergency Management Science and Engineering, University of Chinese Academy of Sciences},
            city={Beijing},
            postcode={100049}, 
            country={China}}
        

\begin{abstract}

Regional differences in carbon emission efficiency arise from disparities in resource distribution, industrial structure, and development level, which are often influenced by government policy preferences. However, currently, most studies fail to consider the impact of government policy preferences and data uncertainty on carbon emission efficiency. To address the above limitations, this study proposes a hybrid model based on $\delta$-slack-based model ($\delta$-SBM) and ordinal priority approach (OPA) for measuring carbon emission efficiency driven by government policy preferences under data uncertainty. The proposed $\delta$-SBM-OPA model incorporates constraints on the importance of input and output variables under different policy preference scenarios. It then develops the efficiency optimization model with Farrell frontiers and efficiency tapes to deal with the data uncertainty in input and output variables. This study demonstrates the proposed model by analyzing industrial carbon emission efficiency in Chinese provinces in 2021. It examines the carbon emission efficiency and corresponding clustering results of provinces under three types of policies: economic priority, environmental priority, and technological priority, with varying priority preferences. The results indicate that the carbon emission efficiency of the 30 provinces can mainly be categorized into technology-driven, development-balanced, and transition-potential types, with most provinces achieving optimal efficiency under the technology-dominant preferences across all policy scenarios. Ultimately, this study suggests a tailored roadmap and crucial initiatives for different provinces to progressively and systematically work towards achieving the low carbon goal.

\end{abstract}




\end{frontmatter}

\section{Introduction}
\label{sec-01}

The increasing emissions of greenhouse gases, represented by carbon dioxide, are exacerbating global climate change \citep{A.J.P.K.H.S22}. China, the world's largest emitter of carbon dioxide, actively engages in international climate cooperation, taking responsibility for global emission reduction amid significant pressure to cut emissions and conserve energy. At the 75th session of the United Nations General Assembly, the Chinese government pledged to `strive to peak carbon dioxide emissions before 2030 and achieve carbon neutrality before 2060' (i.e., the dual carbon strategy) \citep{Q.L.J23}. Under the premise of sustained economic growth, how to effectively promote carbon emission efficiency (CEE) has become an essential issue that China needs to address urgently. Currently, China's industrial sector contributes 40.1 $\%$ of GDP, but its energy consumption and carbon emissions account for 67.9 $\%$ and 84.2 $\%$ of the national total, respectively \citep{C.L.L.Z18}. Improving industrial carbon emission efficiency (ICEE) has become crucial to achieving the dual carbon strategy. Notably, there are significant regional differences in industrial sectors in terms of energy usage, operational efficiency, emission reduction potential, and technological levels, which leads to significant heterogeneity in efficiency across regions \citep{L.H.W23}. Therefore, the government must develop region-specific evaluation systems for ICEE based on the regional endowment, which usually reflects the policy preference.

Current research on CEE assessment from a total factor perspective categorizes methods into parametric and non-parametric methods. This study specifically focuses on non-parametric methods due to the practical challenge of obtaining a predetermined production function required by parametric methods \citep{C.D.M.S.X22}. Among non-parametric methods, data envelopment analysis (DEA) and its extensions, such as BCC \citep{L.Z.Z.L23}, CCR \citep{D.Y.W.C19}, SBM \citep{N.F.K.S23}, super-SBM \citep{G.Y.C21}, NDDF \citep{Y.H.R23} (with abbreviations in Table \ref{tab-01}), are widely employed for assessing CEE at national and regional scales. However, few studies have incorporated considerations of government policy preferences and uncertainties in input data introduced by human factors into the assessment frameworks of CEE \citep{Y.J.W.J.B24}. To overcome the above limitations, this study proposes an integrated approach based on $\delta$-slack-based model ($\delta$-SBM) and ordinal priority approach (OPA) for CEE analysis. This study examines the efficiency differences under economic, environmental, and technological priority policies and their preference scenario through an illustrative demonstration of the ICEE across 30 provinces of China in 2021. Then, K-means clustering is employed to analyze the weight frontiers of input and output variables across provinces under various policy scenarios, identifying groups with similar characteristics. Benchmark provinces in each category are identified by comparing the optimal efficiencies of provinces in each group. Finally, this study offers policy recommendations for different provinces to achieve carbon emission reduction, offering practical guidance for industrial sectors.

The primary contribution of this study lies in proposing $\delta$-SBM-OPA for analyzing CEE that can address policy preferences and data uncertainties within the decision-making scenario. Specifically:
\begin{itemize}
    \item Methodologically, the proposed model formulates policy preferences as scenario constraints, thereby determining the efficiency of DMUs under the specific policy preference scenarios based on human judgment and actual input and output data. Furthermore, the proposed model establishes efficiency tapes for input and output variables facing data uncertainty and calculate corresponding sensitivity indicators to better distinguish the efficiency of individual DMUs, providing more reliable analytical results.
    \item Practically, the proposed model offers decision-makers a customizable and robust tool for analyzing CEE. Decision-makers can observe changes in the CEE of DMUs under different policy scenarios by setting various policies with specific preferences. Conducting cluster analysis on optimal efficiencies and the weight frontier of input and output variables under different scenarios can reveal the optimal scenarios and developmental paths for DMUs to formulate relevant policies.
\end{itemize}

The remainder of this paper is organized as follows: Section \ref{sec-02} presents the literature review. Section \ref{sec-03} introduces the preliminary related to OPA and $\delta$-SBM. Section \ref{sec-04} proposes the $\delta$-SBM-OPA model. Section \ref{sec-05} demonstrates the proposed model with a case study analyzing ICEE among Chinese provinces in 2021, considering specific policy preferences. Section \ref{sec-06} further discusses the variations in ICEE among Chinese provinces under different policy preferences and offers corresponding policy recommendations. Finally, Section \ref{sec-07} summarizes the findings and future directions.

\section{Literature Review}
\label{sec-02}

CEE is a pivotal metric for assessing the carbon emissions of the industrial sector, reflecting the level of low-carbon economic development. Economically, the total output to total input factors ratio is typically employed to assess CEE. CEE was originally a single-factor measure defined by \citet{Y.Y.Y93} as GDP divided by carbon emissions over time. Subsequent research has introduced many additional indicators, including the carbon index, carbon intensity, energy intensity, and emissions per capita per unit of GDP \citep{S.S24}. However, the single-factor approach is inadequate for capturing the multidimensional aspects of CEE. Consequently, a total factor approach has emerged, incorporating labor scale, capital inputs, and energy consumption to provide a comprehensive view of CEE \citep{M.D.Z.R.F23}. Since the introduction of the total factor concept to energy efficiency measurement, it has gained prominence in academia. Methods for analyzing CEE are broadly categorized into parametric and non-parametric methods. However, parametric methods require a predetermined production function, which presents a practical challenge \citep{D.L.G.Z.Q.Z22}. Thus, this study focuses on the non-parametric method, specifically DEA and its extensions, which currently dominates research in this field. Table \ref{tab-01} outlines the critical literature on non-parametric methods for assessing CEE.

Table \ref{tab-01} illustrates that recent studies have evaluated CEE across national, regional, industry, provincial, and municipal levels, considering regular, embedded carbon emissions, water pollution and carbon neutrality, and coordinated governance perspectives. The primary DEA-based non-parametric methods for CEE include radial, non-radial, and directional distance functions. In terms of the radial model, \citet{L.Z.Z.L23} applied the BCC model to assess changes in industrial eco-efficiency across 16 prefecture-level cities in Anhui, China from a static viewpoint. \citet{D.Y.W.C19} performed a comparative analysis of CEE among 30 provinces in China using the CCR and BCC models. However, these conventional radial models overlook the selection of radial direction in efficiency measurement and encounter issues with slack efficiency measurement \citep{T.M24}. To address these issues, several studies have utilized the non-radial super-SBM model to measure CEE. For example, \citet{J.M.Z.G.H20} applied super-SBM to evaluate CEE in the logistics industry across 12 pilot regions in China. \citet{G.Y.C21} integrated the trade openness factor into the embedded carbon emission perspective and employed super-SBM to analyze CEE across 28 industrial sectors in China. \citet{J.L.W.Y.H.B.X24a} measured the CEE of 30 cities in Northwest China from 2011 to 2020 using a super-efficient SBM model based on the dual perspectives of water pollution and carbon neutrality. \citet{F.F.Y22} utilized super-SBM to assess CEE at 42 thermal power plants in China in 2020 from a microscopic perspective. Meanwhile, to enhance environmental efficiency assessment incorporating undesirable outputs, \citet{C.F.G97} introduced the radial DDF based on Shepherd's approach. However, the radial DDF fails to eliminate inefficiencies caused by input and output slack, potentially leading to overestimating CEE. \citet{F.G10} introduced a generalized NDDF for total factor energy productivity, relaxing the requirement for desired and undesirable outputs to vary proportionally. \citet{F.W09} developed the SBM-DDF model for CEE, which integrates undesirable outputs to mitigate radial and directional biases. Moreover, some studies have proposed a multi-stage DEA model combining parametric and non-parametric approaches \citep{Z.M.S.Y.S22}. Among them, the most representative is the three-stage DEA,which is capable of incorporating environmental factors and random noise in the assessment of DMU efficiency \citep{H.X22}.

However, it is noteworthy that the assessment of CEE depends heavily on the value judgments that policymakers make about the resource allocation scenarios and the future of the economy and the environment \citep{X.W.Z23}. This process highlights the potential impact of policy preferences on CEE, which refers to the specific preferences or prioritized objectives the government holds when formulating policies or selecting options. In the assessment of CEE, policy preferences can influence local behavior and decision-making through a variety of mechanisms that promote the transition of the industrial sector towards a higher level of efficiency and cleaner production patterns \citep{W.K.Z17}. The empirical study conducted by \citet{M.S.W21} revealed significant discrepancies in the carbon emission performance of the manufacturing sector when subjected to scale-oriented and innovation-oriented carbon reduction policy preferences. Therefore, a profound comprehension and rigorous consideration of policy preferences is essential to assess alterations in CEE with precision. Such an analysis will assist the government in formulating more effective carbon emission reduction policies, considering the varying circumstances of different regions. Nevertheless, only a limited number of studies that assess CEE take policy preferences into account. In addition, we should consider the potential implications of data uncertainties, which may arise from factors such as statistical inaccuracies or human interference. These factors could significantly influence the assessment of CEE based on policy preferences, which represents a limitation of the current research \citet{Q.X.J.F.W.J22}. In conclusion, the objective of this study is to propose a model for measuring total factor CEE that can accommodate the various policy preference scenarios and account for potential data uncertainties.

\begin{landscape}
\begin{longtable}{llllll}
\caption{Literature on non-parametric methods for assessing CEE}
\label{tab-01}\\
\hline
Reference & Method                                                                        & Subject                                                                                & Perspective                                                                                                & \begin{tabular}{l} Policy \\ preference \end{tabular}                                                 & \begin{tabular}{l} Data \\ uncertainty \end{tabular}                                                 \\ \hline
\endfirsthead
\multicolumn{6}{c}%
{{\bfseries Table \thetable\ continued from previous page}} \\
\hline
Reference & Method                                                                        & Subject                                                                                & Perspective                                                                                                & \begin{tabular}{l} Policy \\ preference \end{tabular}                                                 & \begin{tabular}{l} Data \\ uncertainty \end{tabular}                                                  \\ \hline
\endhead
\hline
\endfoot
\endlastfoot
    \citet{L.Z.Z.L23}      & BCC                                                                           & 16 cities in Anhui, China                                                      & Regular perspective                                                                                        & \scalebox{0.75}{\usym{2613}} & \scalebox{0.75}{\usym{2613}} \\
    \citet{D.Y.W.C19}      & CCR \& BCC                                                                    & 30 provinces in China                                                                   & Regular perspective                                                                                        & \scalebox{0.75}{\usym{2613}} & \scalebox{0.75}{\usym{2613}} \\
    \citet{J.M.Z.G.H20}      & Super-SBM                                                                     & \begin{tabular}[c]{@{}l@{}}Logistics sector in\\ China’s 12 pilot regions\end{tabular} & \begin{tabular}[c]{@{}l@{}}Strong transportation \\ strategy perspective\end{tabular}                      & \scalebox{0.75}{\usym{2613}} & \scalebox{0.75}{\usym{2613}} \\
    \citet{G.Y.C21}      & Super-SBM                                                                     & Industrial sectors in China                                                             & Embodied carbon emission                           & \scalebox{0.75}{\usym{2613}} & \scalebox{0.75}{\usym{2613}} \\
    \citet{J.L.W.Y.H.B.X24a}      & super-SBM                                                                     & 30 cities in Northwestern China                                                        & \begin{tabular}[c]{@{}l@{}}Water pollution \\ and carbon neutrality\end{tabular}                           & \scalebox{0.75}{\usym{2613}} & \scalebox{0.75}{\usym{2613}} \\
    \citet{F.F.Y22}      & Super-SBM                                                                     & 42 thermal power plants in China                                                        & Microscopic perspective                                                                                    & \scalebox{0.75}{\usym{2613}} & \scalebox{0.75}{\usym{2613}} \\
    \citet{Y.H.R23}      & NDDF                                                                          & 282 cities in China                                                                   & Coordinated governance                              & \scalebox{0.75}{\usym{2613}} & \scalebox{0.75}{\usym{2613}} \\
    \citet{H.X22}      & Three-stage DEA                                                               & Export trade sector in China                                                            & \begin{tabular}[c]{@{}l@{}}Embodied carbon emission \\ and coordinated governance \end{tabular} & \scalebox{0.75}{\usym{2613}} & \scalebox{0.75}{\usym{2613}} \\
    \citet{M.S.W21}      & \begin{tabular}[c]{@{}l@{}}Modified global \\ meta-frontier NDDF\end{tabular} & Manufacturing industry in China                                                        & Regular perspective                                                                                        & \checkmark                                        & \scalebox{0.75}{\usym{2613}} \\
    \citet{W.K.Z17}      & NDDF                                                                          & 286 cities in China                                                                    & Regular perspective                                                                                        & \checkmark                                        & \scalebox{0.75}{\usym{2613}} \\
    \citet{Q.X.J.F.W.J22}      & Robust DEA                                                                    & 30 provinces in China                                                                   & Regular perspective                                                                                        & \scalebox{0.75}{\usym{2613}} & \checkmark                                        \\
    \citet{G.C.W.L23}      & INDEA                                                                         & 30 provinces in China                                                                   & Regular perspective                                                                                        & \scalebox{0.75}{\usym{2613}} & \checkmark                                        \\
      \citet{D.L.G.Z.Q.Z22}      & Super-SBM                                                                     
        & 32 developed countries                                                                 & Regular perspective                                                                                        & \scalebox{0.75}{\usym{2613}} & \scalebox{0.75}{\usym{2613}} \\
   \citet{W.Z.L22}      & Super-SBM                                                                     & 131 countries                                                                          & Regular perspective                                                                                        & \scalebox{0.75}{\usym{2613}} & \scalebox{0.75}{\usym{2613}} \\
    \citet{W.L.L23}      & Super-SBM                                                                    
          & 139 countries                                                                          & Regular perspective                                                                                        & \scalebox{0.75}{\usym{2613}} & \scalebox{0.75}{\usym{2613}} \\
   \citet{D.Z.L.C.G.H.Q.S22}      & Super-SBM                                                                     & 32 developed countries                                                                 & Regular perspective                                                                                        & \scalebox{0.75}{\usym{2613}} & \scalebox{0.75}{\usym{2613}} \\ \hline
          \multicolumn{6}{l}
          {\begin{tabular}{l}
          \small {Abbreviations: DEA: Data envelopment analysis; BCC: Banker Charnes and Cooper's model
          CCR: Charnes, Cooper, and Rhodes's model;} \\ 
          \small {Super-SBM: Super slack-based measure;
          NDDF: Non-radial directional distance function;
          INDEA: Interval number DEA.}
          \end{tabular}}
\end{longtable}
\end{landscape}

\section{Preliminary}
\label{sec-03}

\subsection{Ordinal Priority Approach}
\label{sec-03-01}

Ordinal Priority Approach (OPA) is considered a forefront MCDM technique \citep{A.M.F.L20}. The method applies across diverse contexts of MCDM, encompassing the determination of weights for experts, criteria, and alternatives in group and individual decision-making \citep{W24a}. The strength of OPA lies in its utilization of more stable and readily accessible ranking data as inputs, thereby obtaining the weights for experts, criteria, and alternatives simultaneously through solving a linear programming model \citep{P.D.G.D.K.P23}. OPA has found extensive application in domains such as supplier selection \citep{W.S.C.S.C.G24}, portfolio selection \citep{M.A.D22}, performance evaluation \citep{M.A.D22a}, and project planning \citep{W24, M.S.D.M24}. In this study, we will utilize OPA to incorporate the contextual constraints of policy preference for $\delta$-SBM, assessing the impact of varying policy preferences on the efficiency of DMUs. Table \ref{tab-02} elaborates on the sets, indexes, variables, and parameters of OPA required to compute criteria weights derived from expert evaluation.

\begin{table}[h]
\centering
\caption{Sets, indexes, variables, and parameters for OPA}
\label{tab-02}
\begin{tabular}{@{}lll@{}}
\toprule
Type                       & Notation     & Definition                                        \\ \midrule
\multirow{2}{*}{Index}     & $k$          & Index of experts $(1,2,\ldots,m)$.                \\
                           & $j$          & Index of criteria $(1,2,\ldots,n)$.               \\
\multirow{2}{*}{Set}       & $K$          & Set of experts $\forall k\in K$.                  \\
                           & $J$          & Set of criteria $\forall j\in J$.                 \\
\multirow{2}{*}{Variable} & $Z$          & Objective function.                             \\
                          & $w_{jk}^{r}$ & Weight of criteria $j$ based on the evaluation of expert $k$. \\
\multirow{2}{*}{Parameter} & ${{s}_{k}}$  & Rank of expert $k$.                           \\
                           & ${{r}_{jk}}$ & Rank of criteria $j$ given by expert $k$. \\ \bottomrule
\end{tabular}
\end{table}

The initial step of OPA involves determining the ranks of experts, considering aspects like domain expertise, professional experience, job titles, and positions. Subsequently, each expert independently assigns ranks to criteria based on their own judgment and preferences. Then, Equation (\ref{eq-01}) is formulated to determine the criteria weights.
\begin{equation}
\begin{aligned}
\underset{\mathbf{w},Z}{\mathop{\max }}\,\text{ } & Z \\ 
\text{s.t. } & {{s}_{k}}{{r}_{jk}}(w_{jk}^{r}-w_{jk}^{r+1})\ge Z \quad && \forall j\in J,k\in K \\ 
 & {{s}_{k}}{{r}_{jk}}(w_{jk}^{r=n})\ge Z \quad && \forall j\in J,k\in K \\ 
 & \sum\limits_{k=1}^{m}{\sum\limits_{j=1}^{n}{w_{jk}^{r}}}=1 \\ 
 & w_{jk}^{r}\ge 0 \quad && \forall j\in J,k\in K
\end{aligned}
\label{eq-01}
\end{equation}

After solving Equation (\ref{eq-01}), the weights of criteria are calculated according to Equation (\ref{eq-02}).
\begin{equation}
{{w}_{j}}=\sum\limits_{k=1}^{m}{w_{jk}^{r}} \quad \forall j \in J
\label{eq-02}
\end{equation}

\subsection{$\delta$-Slack-Based Model}
\label{sec-03-02}

DEA, a non-parametric data analysis method, primarily assesses the performance of DMUs with multiple input and output variables \citep{P.P24}. Traditional DEA models, such as the CCR, BCC, ADD, and SBM, are significantly affected by the number of DMUs and the input and output variables \citep{K20}. As the number of DMUs decreases or input and output variables increase, the discriminative ability of traditional DEA models in evaluating DMU efficiency diminishes, tending to allocate more DMUs to technical efficiency scores. In practice, many input and output data exhibit a certain degree of uncertainty. This uncertainty primarily arises from factors such as information loss, knowledge constraints, and human errors, particularly in the field of carbon emissions data statistics \citep{Y.J.W.J.B24, W.C.G24}. The traditional DEA model fails to address efficiency evaluations when dealing with imprecise input and output data \citep{A.H.L.H.K24}. To overcome these limitations, \citet{K.S.M14} introduced the $\delta$-SBM model. It is a type of robust DEA model, exhibiting higher flexibility by rational adjustments to the Farrell frontier of inputs and outputs. It creates an effective band region that distinguishes the efficiency levels among various DMUs. Therefore, this study mainly focuses on $\delta$-SBM as the main body of the proposed model for evaluating the efficiency of the DMUs with imprecise input and output data under policy preference. The indexes, sets, variables, and parameters of the $\delta$-SBM model are shown in Table \ref{tab-03}.
\begin{table}[h]
\centering
\caption{Sets, indexes, variables, and parameters for $\delta$-SBM}
\label{tab-03}
\begin{tabular}{@{}lll@{}}
\toprule
Type & Notation & Definition \\ \midrule
\multirow{4}{*} {Index} 
&$i$ &Index of DMUs $(1,2,\ldots,n)$. \\&$j$ &
\begin{tabular}[c]{@{}l@{}}Index of input and output variables $(1,2,\ldots ,r,\ldots ,r+s)$, \\ where $r$ is the number of the input variables and $s$ is \\ the number of the output variables.\end{tabular} \\
\multirow{2}{*}{Set} &$I$ &Set of DMUs $\forall i\in I$. \\
&$J$ &Set of input and output variables $\forall j\in J$. \\
\multirow{4}{*}{Variable} 
&${{\lambda }_{i}}$ & \begin{tabular}[c]{@{}l@{}}Multipliers used for computing linear combinations of DMUs' \\ input and output variables.\end{tabular} \\
&$s_{lj}^{-}$ &Slack variable of input variable $j$ of DMU $l$. \\
&$s_{lj}^{+}$ &Slack variable of output variable $j$ of DMU $l$. \\
\multirow{10}{*}{Parameter} 
&${x}_{ij}$ &Value of input variable $j$ of DMU $l$. \\
&${y}_{ij}$ &Value of output variable $j$ of DMU $l$. \\
&$w_{j}^{-}$ &Assigned weight of input variable $j$ of DMU $l$. \\
&$w_{j}^{+}$ &Assigned weight of output variable $j$ of DMU $l$. \\
&$\varepsilon $ &\begin{tabular}[c]{@{}l@{}}The degree of freedom to create effective tapes by shifting \\ the input and output of Farrell frontier down/up.\end{tabular} \\
&$\varepsilon _{j}^{-}$ &\begin{tabular}[c]{@{}l@{}}Allowed error with $\varepsilon $ degree of freedom of input variable $j$  \\ of DMU $l$ where $\varepsilon _{j}^{-}=\varepsilon {{x}_{lj}}$.\end{tabular} \\
&$\varepsilon _{j}^{+}$ &
  \begin{tabular}[c]{@{}l@{}}Allowed error with $\varepsilon $ degree of freedom of output variable $j$ \\ of DMU $l$ where $\varepsilon _{j}^{+}=\varepsilon {{y}_{lj}}$.\end{tabular} \\ \bottomrule
\end{tabular}
\end{table}

Given the definition of the notations, the $\delta$-SBM formulation for evaluating the performance of each DMU $l\in I$ is presented in Equation (\ref{eq-03}).
\begin{equation}
\begin{aligned}
\underset{{\mathbf{\lambda }},\mathbf{s}^{-},\mathbf{s}^{+}} {\mathop{\max }}\,\text{ } & \sum\limits_{j=1}^{r}{w_{j}^{-}s_{lj}^{-}}+\sum\limits_{j=r+1}^{r+s}{w_{j}^{+}s_{lj}^{+}} \\ 
 \text{s.t. } & \sum\limits_{i=1}^{n}{{{\lambda }_{i}}{{x}_{ij}}}+s_{lj}^{-}={{x}_{lj}}+\varepsilon _{j}^{-} \quad && \forall j\in [r] \\ 
 & \sum\limits_{i=1}^{n}{{{\lambda }_{i}}{{y}_{ij}}}-s_{lj}^{+}={{y}_{lj}}+\varepsilon _{j}^{+} \quad && \forall j\in [s] \\ 
 & {{x}_{lj}}-s_{lj}^{-}\ge 0 \quad && \forall j\in [r] \\ 
 & {{y}_{lj}}+s_{lj}^{+}-2\varepsilon _{j}^{+}\ge 0 \quad && \forall j\in [s] \\ 
 & {{\lambda }_{i}},s_{lj}^{-},s_{lj}^{+}\ge 0 \quad && \forall i\in [n], j\in [s]\\ 
\end{aligned}
\label{eq-03}
\end{equation}

After solving Equation (\ref{eq-03}), the best technical efficient target and score of DMU $l$ with $\varepsilon $ degree of freedom can be represented as Equations (\ref{eq-04}) and (\ref{eq-05}).
\begin{equation}
\left\{ \begin{matrix}
   x_{lj}^{*}={{x}_{lj}}-s{{_{lj}^{-}}^{*}}+\varepsilon _{j}^{-}  \\
   y_{lj}^{*}={{y}_{lj}}+s{{_{lj}^{+}}^{*}}-\varepsilon _{j}^{+}  \\
\end{matrix} \right.
\label{eq-04}
\end{equation}
\begin{equation}
{{\gamma}_{l}}=\frac{{\sum\limits_{j=r+1}^{r+s}{w_{j}^{+}{{y}_{lj}}}} \bigg/ {\sum\limits_{j=1}^{r}{w_{j}^{-}{{x}_{lj}}}}\;}{{\sum\limits_{j=r+1}^{r+s}{w_{j}^{+}y_{lj}^{*}}} \bigg/ {\sum\limits_{j=1}^{r}{w_{j}^{-}x_{lj}^{*}}}\;}
\label{eq-05}
\end{equation}

The lower and upper bound of efficient target of DMU $l$ with $\varepsilon $ degree of freedom be represented as Equations (\ref{eq-06}) and (\ref{eq-07}), respectively.
\begin{equation}
\left\{ \begin{matrix}
   x_{lj}^{+}={{x}_{lj}}-s{{_{lj}^{-}}^{*}}+2\varepsilon _{j}^{-}  \\
   y_{lj}^{+}={{y}_{lj}}+s{{_{lj}^{+}}^{*}}-2\varepsilon _{j}^{+}  \\
\end{matrix} \right.
\label{eq-06}
\end{equation}
\begin{equation}
\left\{ \begin{matrix}
   x_{lj}^{-}={{x}_{lj}}-s{{_{lj}^{-}}^{*}}  \\
   y_{lj}^{-}={{y}_{lj}}+s{{_{lj}^{+}}^{*}}  \\
\end{matrix} \right.
\label{eq-07}
\end{equation}

The sensitivity score of DMU $l$ for the uncertainty efficiency tape with $\varepsilon $ degree of freedom is shown in Equation  (\ref{eq-08}).
\begin{equation}
\eta_{l} = \frac{{\sum\limits_{j=r+1}^{r+s}{w_{j}^{+}y_{lj}^{+}}} \bigg/{\sum\limits_{j=1}^{r}{w_{j}^{-}x_{lj}^{+}}}\;}{{\sum\limits_{j=r+1}^{r+s}{w_{j}^{+}y_{lj}^{-}}} \bigg/ {\sum\limits_{j=1}^{r}{w_{j}^{-}x_{lj}^{-}}}\;}
\label{eq-08}
\end{equation}

Notably, the assigned weight $w_{j}^{-}$ of each input variable can be defined as $1/{\underset{i}{\mathop{\min}} \{{{x}_{ij}}\}}$, $1 / {\underset{i}{\mathop{\max }} \{{{x}_{ij}}\}}$ or $1 / {\underset{i}{\mathrm{avg}} \{{x}_{ij}\}}$ and so on. And the weights $w_{j}^{+}$ of each output variables can be defined in the same way of ${y_{ij}}$. However, when setting weights for input and output variables, it essentially involves non-dimensional standardization in the original $\delta$-SBM model, overlooking the subjective judgment and preference of decision-makers \citep{G.C23, M.A.D22a}. This factor might lead to impractical solutions, particularly when considering impact of varying policy preferences on CEE analysis. Hence, it becomes necessary to incorporate weights $w_{j}^{-}$ and $w_{j}^{+}$ as variables of the $\delta$-SBM model, originating from the decision-makers' preference perspective. In the next section, we will propose the $\delta$-SBM-OPA model for carbon emission analysis under policy preference with the aid of Equations (\ref{eq-01}) and (\ref{eq-03}).

\section{The Hybrid $\delta$-SBM-OPA model for Carbon Emission Efficiency Analysis under Policy Preference}
\label{sec-04}

In this section, a hybrid $\delta$-SBM-OPA model is proposed to analyze the CEE under multiple scenarios with different government policy preferences. The first step is to derive the dual problem of the original $\delta$-SBM model since it offers lucid guidance on the weightage information to criteria (i.e., input and output variables). These weights delineate how each DMU prioritizes its input and output variables (e.g., capital inputs, industrial output, and carbon emissions) when striving for optimal efficiency of carbon emission considering policy preference. Moreover, the transformation of a dual problem converts the initial nonlinear optimization problem into a linear optimization and reduces the number of decision variables involved. The dual problem of the original $\delta$-SBM model is shown in Equation (\ref{eq-09}).
\begin{equation}
\begin{aligned}
\underset{\mathbf{v}^{-},\mathbf{u}^{+},\mathbf{\theta}^{-},\mathbf{\sigma}^{+}}{\mathop{\min }}\,\text{ } & \sum\limits_{j=1}^{r}{({{x}_{lj}}+\varepsilon _{j}^{-})v_{j}^{-}}+\sum\limits_{j=1}^{r}{{{x}_{lj}}\theta _{j}^{-}} \\ 
&\quad -\sum\limits_{j=r+1}^{r+s}{({{y}_{lj}}+\varepsilon _{j}^{+})u_{j}^{+}}-\sum\limits_{j=r+1}^{r+s}{({{y}_{lj}}-2\varepsilon _{j}^{+})\sigma _{j}^{+}} \\ 
\text{s.t. } & \sum\limits_{j=1}^{r}{{{x}_{ij}}v_{j}^{-}}-\sum\limits_{j=r+1}^{r+s}{{{y}_{ij}}u_{j}^{+}}\ge 0 \quad && \forall i\in [n] \\ 
& v_{j}^{-}+\theta _{j}^{-}\ge w_{j}^{-} \quad && \forall j\in [r] \\ 
& u_{j}^{+}+\sigma _{j}^{+}\ge w_{j}^{+} \quad && \forall j\in [s] \\ 
& v_{j}^{-},\theta _{j}^{-},u_{j}^{+}\ge 0,\sigma _{j}^{+}\le 0 \quad && \forall j\in [r+s]
\end{aligned}
\label{eq-09}
\end{equation}

Equations (\ref{eq-01}) and (\ref{eq-09}) are integrated into a multi-objective optimization model, illustrated in Equation (\ref{eq-10}), to account for the influence of policy preferences on the importance of input and output variables in analyzing CEE.
\begin{equation}
\begin{aligned}
\underset{\begin{array}{c} \mathbf{v}^{-},\mathbf{u}^{+},\mathbf{\theta}^{-},\mathbf{\sigma}^{+}, \\ \mathbf{w}, Z
\end{array}}{\min} \text{ } & \sum\limits_{j=1}^{r}{({{x}_{lj}}+\varepsilon _{j}^{-})v_{j}^{-}}+\sum\limits_{j=1}^{r}{{{x}_{lj}}\theta _{j}^{-}} \\
& \quad  -\sum\limits_{j=r+1}^{r+s}{({{y}_{lj}}+\varepsilon _{j}^{+})u_{j}^{+}}-\sum\limits_{j=r+1}^{r+s}{({{y}_{lj}}-2\varepsilon _{j}^{+})\sigma _{j}^{+}} \\ 
\underset{\mathbf{w}, Z}{\max} \text{ } & Z \\ 
\text{s.t. } & \sum\limits_{j=1}^{r}{{{x}_{ij}}v_{j}^{-}}-\sum\limits_{j=r+1}^{r+s}{{{y}_{ij}}u_{j}^{+}}\ge 0 \quad && \forall i\in [n] \\ 
& v_{j}^{-}+\theta _{j}^{-} \ge {\sum\limits_{k=1}^{p}{w_{jk}^{t}}} \bigg/ {\underset{i}{\mathop{\max }}\,\{{{x}_{ij}}\}} \quad && \forall j\in [r] \\ 
& u_{j}^{+}+\sigma _{j}^{+} \ge {\sum\limits_{k=1}^{p}{w_{jk}^{t}}} \bigg/ {\underset{i}{\mathop{\min }}\,\{{{y}_{ij}}\}} \quad && \forall j\in [s] \\ 
& {{t}_{k}}{{t}_{jk}}(w_{jk}^{t}-w_{jk}^{t+1})\ge Z \quad && \forall j\in [s]+[r],  k \in [p] \\ 
 & {{t}_{k}}{{t}_{jk}}(w_{jk}^{t=r+s})\ge Z \quad && \forall j\in [s]+[r],  k \in [p] \\ 
& \sum\limits_{k=1}^{p}{\sum\limits_{j=1}^{r+s}{w_{jk}^{t}}}=1 \\ 
& v_{j}^{-},\theta _{j}^{-},u_{j}^{+},w_{jk}^{t}\ge 0,\sigma _{j}^{+}\le 0 \quad && \forall j\in [s]+[r],  k \in [p]
\end{aligned}
\label{eq-10}
\end{equation}

Where $\varepsilon _{j}^{+} = 1 / \max_{i}{\underset{i}{\mathop{\max }}\,\{{{x}_{ij}}\}}$ and $\varepsilon _{j}^{-} = 1 / {\underset{i}{\mathop{\min }}\,\{{{y}_{ij}}\}}$. In Equation (\ref{eq-10}), the first objective function and the first constraint belong to the dual problem of $\delta$-SBM in Equation (\ref{eq-09}). The second objective function and the fourth, fifth, and sixth constraints belong to OPA in Equation (\ref{eq-01}). The right-hand side of the second and third constraint is the output of OPA in Equation (\ref{eq-01}), and the left-hand side is the input of  $\delta$-SBM in Equation (\ref{eq-09}). From a modeling perspective, OPA provides $\delta$-SBM with lower bound constraints on the importance of input and output variables that take policy preferences into account.

Equation (\ref{eq-10}) presents a multi-objective optimization model, which can be solved through various methods like Pareto-optimality, goal programming, budgeted-constraint approach, and the max-min approach \citep{M.A.D22a}. This study utilizes the weighted max-min approach due to its flexibility for decision-makers to express the relative importance between $\delta$-SBM and OPA. Since the scales differ between $\delta$-SBM and OPA, Equation (\ref{eq-11}) is utilized to transform the objective functions into non-dimensional counterparts, with values fall within the range of [0,1].
\begin{equation}
f_{k}^{trans}=\frac{\max \{{{f}_{k}}(x)\}-{{f}_{k}}(x)}{\max \{{{f}_{k}}(x)\}-\min \{{{f}_{k}}(x)\}} \quad \forall k\in [n+1]
\label{eq-11}
\end{equation}

\begin{lem}
The optimal value of the objective function $Z^{*} = \max Z$ in Equation (\ref{eq-01}) is confined within the interval $[0,1]$.
\label{lemma-01}
\end{lem}

\newproof{lemma-proof-01}{Proof of Lemma \ref{lemma-01}}
\begin{lemma-proof-01}
(1) Show that the lower bound of the optimal value $Z^{*}$ is 0. By the derivation of OPA, the attribute that has a higher ranking $r+1$ is dominated by the one with a lower ranking $r$, i.e., $A_{jk}^r \succeq A_{jk}^{r+1}$, which is equivalent to $w_{jk}^{r} \ge w_{jk}^{r+1}$. For $\forall j \in J, k \in K$, we have $s_{k}r_{jk}(w_{jk}^r-w_{jk}^{r+1}) \ge 0$ and $s_{k}r_{jk}(w_{jk}^{r=n}) \ge 0$. The objective function is to maximize $Z$, then there exists $\underset{j,k}{\min}\left\{s_{k}r_{jk}(w_{jk}^r-w_{jk}^{r+1})\right\} = \max Z = Z^{*} \ge 0$ or $\underset{j,k}{\min}\left\{s_{k}r_{jk}(w_{jk}^{r=n})\right\} = \max Z = Z^{*} \ge 0$ such that $\max Z = Z^{*} \ge 0$ holds. Thus, the lower bound of the optimal value $Z^{*}$ is 0.

(2) Show that the upper bound of the optimal value $Z^{*}$ is 1. Suppose that there exists $\epsilon_{jk}$ such that $s_{k}r_{jk}(w_{jk}^r-w_{jk}^{r+1}) = Z^{*} + \epsilon_{jk}$ and $s_{k}r_{jk}(w_{jk}^{r=n}) = Z^{*} + \epsilon_{jk}$ for $\forall j \in J, k \in K$. If $\epsilon_{jk} < 0$, we have $s_{k}r_{jk}(w_{jk}^r-w_{jk}^{r+1}) = \Bar{Z} < Z^{*}$ or $s_{k}r_{jk}(w_{jk}^{r=n}) = \Bar{Z} < Z^{*}$, which contradicts the objective of maximizing the minimum $Z$. Thus, $\epsilon_{jk} \ge 0$ and there must be at least one $\epsilon_{jk} = 0$ such that $s_{k}r_{jk}(w_{jk}^r-w_{jk}^{r+1}) = Z^{*}$ and $s_{k}r_{jk}(w_{jk}^{r=n}) = Z^{*}$. Then, the cumulative sum of the last $k$ constraints for each expert $j \in J$ in ascending order yields
\[
w_{jk}^r = \frac{1}{s_{k}} \left( \sum_{h=r_{jk}}^{n} \frac{1}{h} \right) (Z^{*} + \epsilon_{jk}).
\]
Substituting the normalized constraint, we have 
\[
\begin{aligned}
    \sum_{k = 1}^{m} \sum_{j=1}^{n} w_{jk}^{r} = 1 & \Leftrightarrow Z^{*}\sum_{k = 1}^{m} \sum_{j=1}^{n} \left(\frac{1}{s_{k}} \sum_{h=r_{jk}}^{n} \frac{1}{h} \right) + \sum_{k = 1}^{m} \sum_{j=1}^{n} \left( \frac{\epsilon_{jk}}{s_{k}}  \sum_{h=r_{jk}}^{n} \frac{1}{h} \right) = 1, \\
    & \Leftrightarrow Z^{*} = \underbrace{\left( 1-\sum_{k = 1}^{m} \sum_{j=1}^{n} \left( \frac{\epsilon_{jk}}{s_{k}}  \sum_{h=r_{jk}}^{n} \frac{1}{h} \right) \right)}_{\leq 1} \Bigg/ \underbrace{\left( n \sum_{k = 1}^{m} \frac{1}{k} \right)}_{\ge1}.
\end{aligned}
\]
It follows that $Z^{*} \leq 1$, which implies the upper bound of the optimal value $Z^{*}$ is 1.
\qed
\end{lemma-proof-01}

Lemma \ref{lemma-01} implies that the optimal value $Z$ is dimensionless and lies within the interval [0,1]. Given its suitable numerical scale, further transformation in Equation (\ref{eq-11}) is unnecessary.

Denote ${{U}_{S}}$ and ${{U}_{P}}$ as the weights of the objective functions of the $\delta$-SBM and OPA models, respectively, where ${{U}_{S}}+{{U}_{P}}=1$. Then, Equation (\ref{eq-10}) can be transferred into weighted max-min form, as shown in Equation (\ref{eq-12}).
\begin{equation}
\begin{aligned}
\underset{\mathbf{v}^{-},\mathbf{u}^{+},\mathbf{\theta}^{-},\mathbf{\sigma}^{+}, \mathbf{w}, Z}{\max \min} \text{ } & \{{{U}_{S}}[f_{k}^{trans}(\sum\limits_{j=1}^{r}{({{x}_{lj}}+\varepsilon _{j}^{-})v_{j}^{-}}+\sum\limits_{j=1}^{r}{{{x}_{lj}}\theta _{j}^{-}} \\ 
 & \quad -\sum\limits_{j=r+1}^{r+s}{({{y}_{lj}}+\varepsilon _{j}^{+})u_{j}^{+}}-\sum\limits_{j=r+1}^{r+s}{({{y}_{lj}}-2\varepsilon _{j}^{+})\sigma _{j}^{+}})], {{U}_{P}}Z \} \\ 
\text{s.t. } & \sum\limits_{j=1}^{r}{{{x}_{ij}}v_{j}^{-}}-\sum\limits_{j=r+1}^{r+s}{{{y}_{ij}}u_{j}^{+}}\ge 0 \quad && \forall i\in [n] \\ 
& v_{j}^{-}+\theta _{j}^{-} \ge {\sum\limits_{k=1}^{p}{w_{jk}^{t}}} \bigg/ {\underset{i}{\mathop{\max }}\,\{{{x}_{ij}}\}} \quad && \forall j\in [r] \\ 
& u_{j}^{+}+\sigma _{j}^{+} \ge {\sum\limits_{k=1}^{p}{w_{jk}^{t}}} \bigg/ {\underset{i}{\mathop{\min }}\,\{{{y}_{ij}}\}} \quad && \forall j\in [s] \\ 
& {{t}_{k}}{{t}_{jk}}(w_{jk}^{t}-w_{jk}^{t+1})\ge Z \quad && \forall j\in [s]+[r],  k \in [p] \\ 
 & {{t}_{k}}{{t}_{jk}}(w_{jk}^{t=r+s})\ge Z \quad && \forall j\in [s]+[r],  k \in [p] \\ 
& \sum\limits_{k=1}^{p}{\sum\limits_{j=1}^{r+s}{w_{jk}^{t}}}=1 \\ 
& v_{j}^{-},\theta _{j}^{-},u_{j}^{+},w_{jk}^{t}\ge 0,\sigma _{j}^{+}\le 0 \quad && \forall j\in [s]+[r],  k \in [p]
\end{aligned}
\label{eq-12}
\end{equation}

The optimization model in max-min form can be further transformed into a linear programming problem by variable substitution. Let
\begin{equation}
\begin{aligned}
 \xi = \min \text{ } & \{{{U}_{S}}[f_{k}^{trans}(\sum\limits_{j=1}^{r}{({{x}_{lj}}+\varepsilon _{j}^{-})v_{j}^{-}}+\sum\limits_{j=1}^{r}{{{x}_{lj}}\theta _{j}^{-}} \\ 
 & \quad -\sum\limits_{j=r+1}^{r+s}{({{y}_{lj}}+\varepsilon _{j}^{+})u_{j}^{+}}-\sum\limits_{j=r+1}^{r+s}{({{y}_{lj}}-2\varepsilon _{j}^{+})\sigma _{j}^{+}})], {{U}_{P}}Z\}. 
\end{aligned}
\label{eq-13}
\end{equation}

Then, substituting Equation (\ref{eq-13}) into Equation (\ref{eq-12}) yields a single objective linear optimization model, as demonstrated by Proposition \ref{pro-01}.
\begin{pro} 
Given the input and output values of all DMUs, along with the variable prioritization based on a particular policy preference scenario, the $\delta$-SBM-OPA model for CEE assessment considering policy preference is formulated as Equation (\ref{eq-14}).
\begin{equation}
\begin{aligned}
\underset{\begin{array}{c} \mathbf{v}^{-},\mathbf{u}^{+},\mathbf{\theta}^{-}, \\ \mathbf{\sigma}^{+},\mathbf{w}, Z, \xi 
\end{array}}{\max} \text{ } & \xi  \\ 
\mathrm{s.t.}\text{ } & {{U}_{S}}[f_{k}^{trans}(\sum\limits_{j=1}^{r}{({{x}_{lj}}+\varepsilon _{j}^{-})v_{j}^{-}}+\sum\limits_{j=1}^{r}{{{x}_{lj}}\theta _{j}^{-}} \\
& \quad -\sum\limits_{j=r+1}^{r+s}{({{y}_{lj}}+\varepsilon _{j}^{+})u_{j}^{+}}-\sum\limits_{j=r+1}^{r+s}{({{y}_{lj}}-2\varepsilon _{j}^{+})\sigma _{j}^{+}})]-\xi \ge 0 \\ 
& {{U}_{P}}Z-\xi \ge 0 \\
& \sum\limits_{j=1}^{r}{{{x}_{ij}}v_{j}^{-}}-\sum\limits_{j=r+1}^{r+s}{{{y}_{ij}}u_{j}^{+}}\ge 0 && \forall i\in [n] \\ 
& v_{j}^{-}+\theta _{j}^{-} \ge {\sum\limits_{k=1}^{p}{w_{jk}^{t}}} \bigg/ {\underset{i}{\mathop{\max }}\,\{{{x}_{ij}}\}}  && \forall j\in [r] \\ 
& u_{j}^{+}+\sigma _{j}^{+} \ge {\sum\limits_{k=1}^{p}{w_{jk}^{t}}} \bigg/ {\underset{i}{\mathop{\min }}\,\{{{y}_{ij}}\}}  && \forall j\in [s] \\ 
& {{t}_{k}}{{t}_{jk}}(w_{jk}^{t}-w_{jk}^{t+1})\ge Z  && \forall j\in [s]+[r],  k \in [p] \\ 
 & {{t}_{k}}{{t}_{jk}}(w_{jk}^{t=r+s})\ge Z  && \forall j\in [s]+[r],  k \in [p] \\ 
& \sum\limits_{k=1}^{p}{\sum\limits_{j=1}^{r+s}{w_{jk}^{t}}}=1 \\ 
& v_{j}^{-},\theta _{j}^{-},u_{j}^{+},w_{jk}^{t}\ge 0,\sigma _{j}^{+}\le 0  && \forall j\in [s]+[r],  k \in [p]
\end{aligned}
\label{eq-14}
\end{equation}
\label{pro-01}
\end{pro}
The optimal solution set $(v{{_{j}^{-}}^{*}},\theta {{_{j}^{-}}^{*}},u{{_{j}^{+}}^{*}},\sigma {{_{j}^{+}}^{*}},w_{jk}^{*})$ from Proposition \ref{pro-01} using the max-min method may not all be efficient. Nevertheless, at least one element of this set is an efficient point for the multi-objective optimization model described in Equation (\ref{eq-10}) \citep{W24a, B.C24}. After solving Equation (\ref{eq-14}), Equations (\ref{eq-04})-(\ref{eq-08}) can be employed to calculated the best technical efficiency score, lower and upper bound of efficiency, and sensitivity score of DMD $l$ under specific policy preference. Ultimately, decision-makers can cluster DMUs based on the weights of input and output variables, facilitating an analysis of consolidating efficiency targets within each category. This approach helps to develop the most effective developmental path for DMUs under specific policy preference.

\section{Illustrative Demonstration of Industrial Sector in Chinese Provinces}
\label{sec-05}

\subsection{Data Collection}
\label{sec-05-01}

This study applies the proposed $\delta$-SBM-OPA model to analyzing ICEE of 30 provinces in China in 2021. This study select capital, labor, energy, and technology as input variables, with industry output and CO2 emissions as output variables \citep{W.Z.L22}. Compared to the common input and output variables utilized in other ICEE studies, this study introduces technology as a novel input variable, thereby facilitating a more comprehensive ICEE analysis. As technology develops, the incorporation of the technology factor into production processes and energy consumption has the potential to significantly impact carbon emissions. The following presents the data collection and processing of input and output variables.

Input variable. Considering the influence of capital input characteristics on industrial sector output, this research focuses on capital stock ($K$), specifically using net fixed assets of large industrial enterprises as a proxy \citep{Z.Y.W.X.L.Z.Z.W.H23}. Unlike using the perpetual inventory method, this study avoids assumptions about depreciation rates, which are often arbitrarily set around figures like9.6$\%$ or 6$\%$. Labor ($L$) represents the average number of employees in industrial enterprises above the designated size \citep{Z22}. About 95$\%$ of the carbon dioxide produced by human activities comes from the use of fossil fuels. Energy ($E$) is therefore represented by the final consumption of the eight primary fossil fuels, converted into standard coal equivalent. These fuels includes hard coal, coke, crude oil, petrol, kerosene, diesel, heating oil, and natural gas. Internal expenditure on R$\&$D by industrial enterprises above the designated size represents technology ($T$) \citep{D.Z.L.C.G.H.Q.S22}.

Output variable. The primary business income ($Y$) of industrial enterprises above the designated size represents industrial output. Notably, most studies do not use gross industrial output value as a measure of industrial output value, mainly because the Chinese Industrial Economy Statistical Yearbook stopped reporting data on gross industrial output value in 2012 \citep{M.S.W21}. For carbon dioxide (CO2) emissions ($C$) , given that there is no direct access to industrial CO2 emissions by region from any statistical review or database. This study uses the method of the International Panel on Climate Change to estimate industrial CO2 emissions across 30 provinces of China in 2021 \citep{W.L.L23}. The formula for measuring carbon dioxide emissions $CE$ follows:
\begin{equation}
    CE = \sum\limits_{i=1}^{8}CE_{i} = \sum\limits_{i=1}^{8} E_{i} \times NCV_{i} \times CEF_{i} \times COF_{i} \times 44/12,
\end{equation}
where $i$ is the index of fossil fuel type, and $CE_{i}$, ${E_{i}}$, $NCV_{i}$, $CEF_{i}$, and $COF_{i}$ represents the carbon dioxide emissions, consumption, average lower heating value, carbon content per unit calorific value, and carbon oxidation rate of fossile fuel $i$, respectively. This study adjusts the carbon emission factors according to the National Development and Reform Commission .

This study uses data from the 2021 China Industrial Statistical Yearbook, the China Energy Statistical Yearbook, the China Provincial Statistical Yearbook, and the China Science and Technology Statistical Yearbook to analyze ICEE in 30 Chinese provinces. It should be noted that the data related to Tibet, Hong Kong, Macao, and Taiwan are not discussed in this paper, as some of the data for these regions are missing. Table \ref{tab-01} shows the descriptive statistics of the data . 

\begin{table}[h]
\centering
\caption{Descriptive statistics of input and output variables}
\label{tab-04}
\begin{tabular}{@{}lllllll@{}}
\toprule
Index  & Unit            & Observations & Min     & Max      & Mean     & Std.Dev  \\ \midrule
$L$ & $10^4$ persons     & 30           & 11.52   & 1354.17  & 264.94   & 295.49   \\
$K$ & 100 million RMB    & 30           & 1391    & 35789.8  & 12619.23 & 8347.02  \\
$T$ & 100 million RMB    & 30           & 13.85   & 2902.19  & 583.73   & 727.05   \\
$E$ & $10^6$ tons        & 30           & 0.62    & 113.43   & 32.84    & 24.62    \\
$Y$ & 100 million RMB    & 30           & 2676.14 & 173649.70 & 43804.72 & 41142.54 \\
$C$ & $10^6$ tons        & 30           & 1.35    & 323.75   & 90.16    & 69.59    \\ \bottomrule
\end{tabular}
\end{table}

\subsection{Policy Preference Analysis and Setting}
\label{sec-05-02}

Policy preference refers to the government tendency to focus on particular objective or interests when prioritizing policies \citep{Y.J.W.J.B24}. This tendency is essential in guiding government decisions on resource allocation, policy implementation, and monitoring \citep{Y.J.W.J.B24}. Policy preferences also reflect the degree of integration and importance governments attach to individual national development strategies in policy formulation. Studies have shown that the policy preferences of government can significantly impact carbon emissions \citep{C.D.M.S.X22}. Changes in these preferences, along with shifts in industrial structure and government interventions, profoundly affect the efficacy of reducing carbon emissions. Policy preferences directly shape the prioritization of policy implementation. The significance of input and output variables varies when assessing CEE under different policies. Due to diverse resource endowments, industrial divisions of labor, and developmental stages, disparities in CEE across provinces are inevitable under varying policy preferences. Therefore, governments must develop a rational evaluation system for CEE that reflects local conditions and aligns with policy preferences.

This study introduces three policy scenarios in response to the recent policy focus of China: economic, environmental, and technology priorities. input and output variables within each scenario are prioritized based on policy characteristics. Certain elements are clearly prioritized under each policy preference, while others remain uncertain. Hence, correlation analysis is employed in this study to rank the importance of these elements within each policy. Specifically, this study utilizes data from the primary element under each policy across 30 provinces as a reference series. Pearson correlation tests are conducted on the data of the other elements, ranking them within each policy based on Pearson coefficients from highest to lowest. Table \ref{tab-05} presents the ranking results of input and output variables under each policy. Subsequently, government policy preferences are formed by ranking the importance of these policies. This section demonstrates the proposed model using the policy preference scenario of `P1>P2>P3' as an illustrative example. Moreover, all other possible policy preference scenarios will be analyzed explicitly in Section \ref{sec-06}.

\begin{table}[h]
\centering
\caption{Ranking of input and output variables under different policies}
\label{tab-05}
\begin{tabular}{@{}lllllll@{}}
\toprule
Policy                             & $L$ & $K$ & $T$ & $E$ & $Y$ & $C$ \\ \midrule
Economic priority policy (P1)        & 4 & 2 & 3 & 5 & 1 & 6 \\
Environmental priority policy (P2)   & 3 & 4 & 6 & 2 & 4 & 1 \\
Technological priority policy (P3)      & 4 & 3 & 1 & 6 & 2 & 5 \\ \bottomrule
\end{tabular}
\end{table}

\subsection{Result Analysis}
\label{sec-05-03}
\subsubsection{Regional Differences in Industrial Carbon Emission Efficiency}
\label{sec-05-03-01}

\begin{figure}[h]
    \centering
    \includegraphics[width=0.6\linewidth]{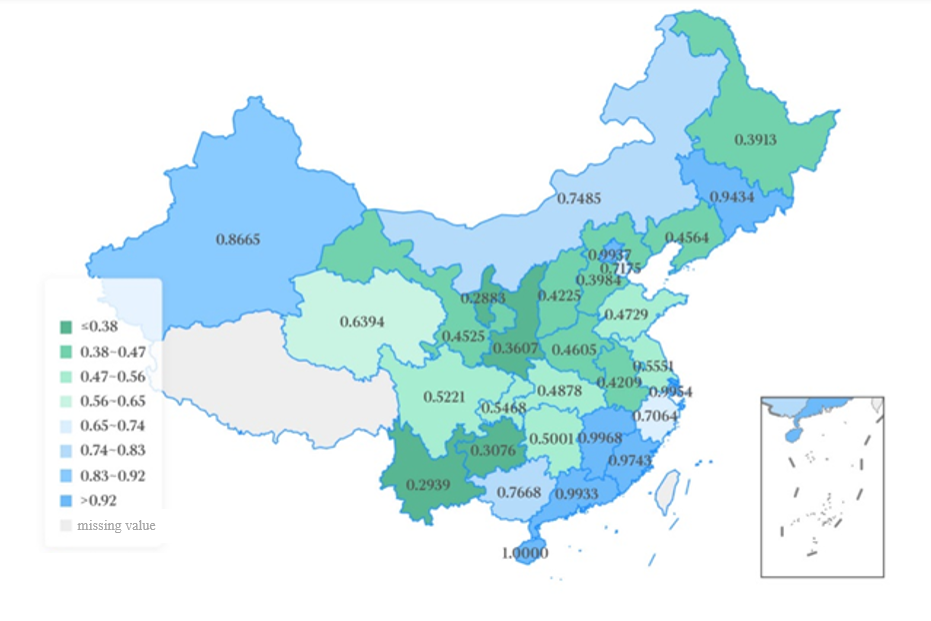}
    \caption{ Industrial Carbon Emission Efficiency Map of 30 Chinese Provinces in 2021}
    \label{fig-01}
\end{figure}

Figure \ref{fig-01} depicts the ICEE across 30 provinces of China in 2021, as calculated by the proposed $\delta$-SBM-OPA method. The results demonstrate that the mean value of ICEE across the 30 provinces is 0.6227, with only 12 provinces exceeding the national average efficiency level. The provinces of Beijing, Shanghai, Guangdong, Jiangxi, and Hainan, respectively, have the highest efficiencies, all exceeding 0.99. The ICEE in Guizhou, Yunnan, Shaanxi, Ningxia, Hebei, Shanxi, Liaoning, Heilongjiang, Anhui, Shandong, Henan, Hubei, and Gansu is notably poor, with values below 0.5. The above indicates that the overall ICEE across China's 30 provinces is low and exhibits considerable variation.

This study categorizes the provinces into eight economic regions divided by the State Council of China. The mean and variance of efficiencies across the provinces involved in the eight regions and the efficiencies for each province are presented in Table\ref{tab-05}. The East and South Coasts exhibit the highest ICEE, with average values of 0.7523 and 0.9892, respectively. The following are the North Coast, Middle Yangtze, Northeast, and Northwest, with the ICEE of 0.6456, 0.6014, 0.5970, and 0.5617, respectively. However, the ICEE of the Southwest and Northwest is notably deficient, with values below 0.5, at 0.4981 and 0.4874, respectively. As for the variance of regional ICEE, the East and South Coasts show quantum differences from the other regions, especially the South Coast at 0.0001. In contrast, the variance in the other regions is within the interval [0.022,0.0607]. As for provincial efficiencies within each region, the results show that all regions except the Southwest, Northwest, and Middle Yellow River regions have at least one province with an efficiency of 0.9 or higher. Noticeably, the three provinces on the South Coast (i.e., Fujian, Guangdong, and Hainan) all have efficiencies of 0.9743 and above. This illustrates that the South Coast performs exceptionally well in terms of ICEE and has the potential to become a national benchmark for ICEE.

\begin{figure}[h]
    \centering
    \includegraphics[width=0.75\linewidth]{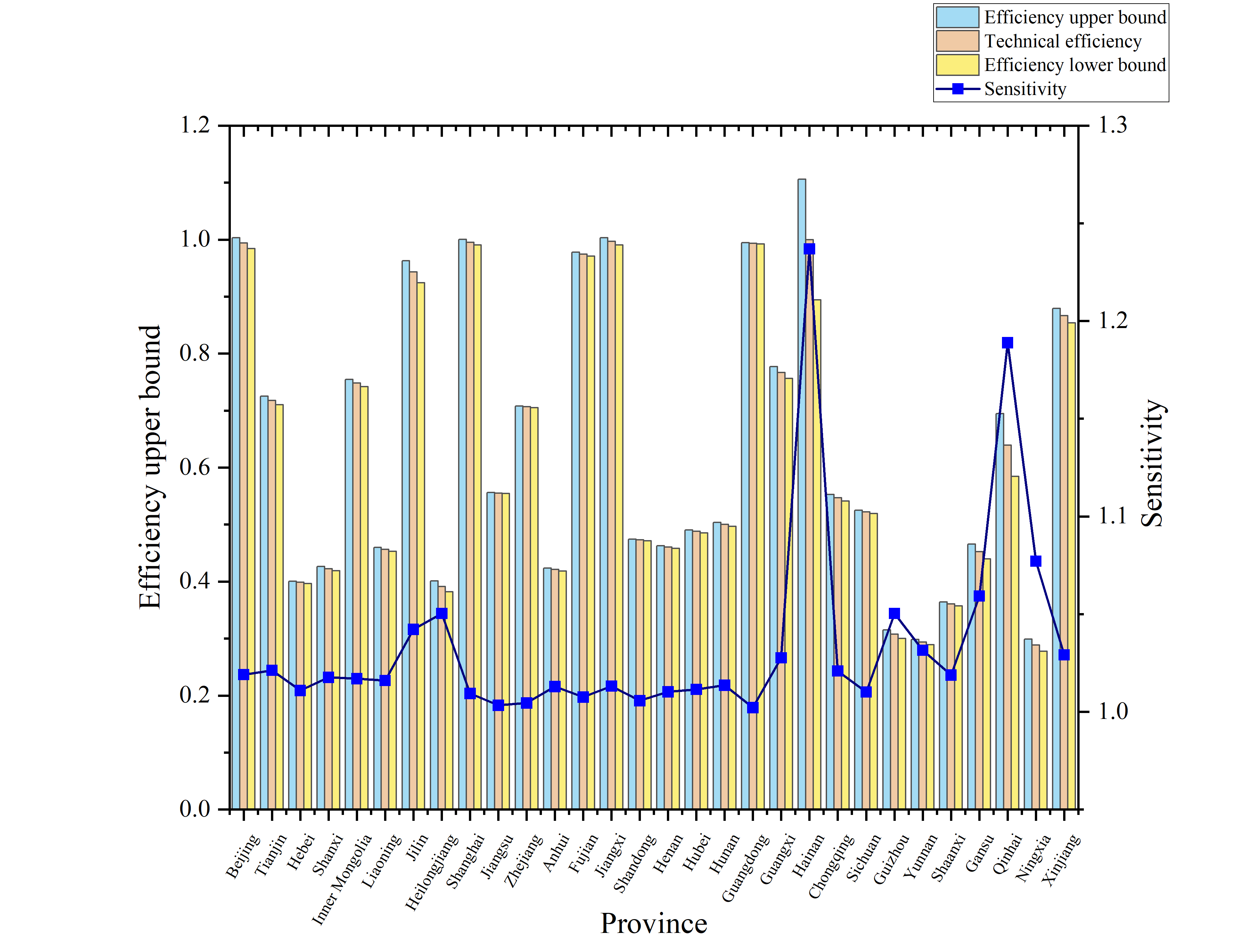}
    \caption{Sensitivity of province carbon emission efficiencies}
    \label{fig-02}
\end{figure}

Subsequently, this study calculates the sensitivity of ICEE for each province within the specified uncertainty interval. Figure \ref{fig-02} depicts the technical efficiency, upper and lower bounds, and sensitivities of ICEE across 30 provinces of China in 2021. The calculations show that the mean and standard deviation of the ICEE sensitivity are 1.0346 and 0.0520, respectively. Notably, Hainan and Qinghai exhibit the highest sensitivity, reaching 1.2369 and 1.1889, respectively. These values exceed the mean plus double the standard deviation, indicating that these two provinces are particularly susceptible to data uncertainty. Even though the ICEE of Hainan has reached the efficiency frontier, its sensitivity score shows that there is still potential for further improvement in its efficiency level. In addition, the sensitivities of Ningxia, Heilongjiang, Guizhou, Gansu, and Jilin are higher than the national average. Heilongjiang, Guizhou, Gansu, and Ningxia also have low ICEE. Overall, evaluating ICEE is not the sole criterion in the context of $\delta$-SBM-OPA. Instead, it is essential to consider the uncertainty-oriented sensitivities of each province comprehensively. Only higher ICEE accompanied by more stable results can be considered efficiency targets.

\begin{landscape}
\begin{table}[]
\caption{Regional characteristics of industrial carbon efficiency}
\label{tab-06}
\begin{tabular}{@{}llllllllll@{}}
\hline
\multicolumn{1}{l}{Region}                        & \multicolumn{1}{l}{Province}  & \multicolumn{1}{l}{\begin{tabular}[l]{@{}l@{}}Efficiency \\ score\end{tabular}} & \multicolumn{1}{l}{\begin{tabular}[l]{@{}l@{}}Mean \\ value\end{tabular}}  & \multicolumn{1}{l}{Variance}                & \multicolumn{1}{l}{Region}                                                                                           & \multicolumn{1}{l}{Province}     & \multicolumn{1}{l}{\begin{tabular}[l]{@{}l@{}}Efficiency \\ score\end{tabular}}  & \multicolumn{1}{l}{\begin{tabular}[l]{@{}lc@{}}Mean \\ value\end{tabular}}              & \multicolumn{1}{l}{Variance}                \\ \midrule
\multirow{4}{*}{North coast}  & Beijing   & 0.9937           & \multirow{4}{*}{0.6456} & \multirow{4}{*}{0.0543} & \multirow{4}{*}{Northwestern}                                                                   & Ningxia      & 0.2883           & \multirow{4}{*}{0.5617} & \multirow{4}{*}{0.0464} \\
                              & Tianjin   & 0.7175           &                         &                         &                                                                                                 & Gansu        & 0.4525           &                         &                         \\
                              & Hebei     & 0.3984           &                         &                         &                                                                                                 & Qinghai      & 0.6394           &                         &                         \\
                              & Shandong  & 0.4729           &                         &                         &                                                                                                 & Xinjiang     & 0.8665           &                         &                         \\
\multirow{3}{*}{East coast}   & Shanghai  & 0.9954           & \multirow{3}{*}{0.7523} & \multirow{3}{*}{0.0034} & \multirow{3}{*}{Northeast}                                                                      & Liaoning     & 0.4564           & \multirow{3}{*}{0.597}  & \multirow{3}{*}{0.0607} \\
                              & Jiangsu   & 0.5551           &                         &                         &                                                                                                 & Jilin        & 0.9434           &                         &                         \\
                              & Zhejiang  & 0.7064           &                         &                         &                                                                                                 & Heilongjiang & 0.3913           &                         &                         \\
\multirow{3}{*}{South coast}  & Fujian    & 0.9743           & \multirow{3}{*}{0.9892} & \multirow{3}{*}{0.0001} & \multirow{4}{*}{\begin{tabular}[c]{@{}l@{}}Middle reaches of \\ the Yellow River\end{tabular}}  & Shanxi       & 0.4225           & \multirow{4}{*}{0.4981} & \multirow{4}{*}{0.0222} \\
                              & Guangdong & 0.9933           &                         &                         &                                                                                                 & Neimenggu    & 0.7485           &                         &                         \\
                              & Hainan    & 1                &                         &                         &                                                                                                 & Henan        & 0.4605           &                         &                         \\
\multirow{5}{*}{Southwestern} & Guangxi   & 0.7668           & \multirow{5}{*}{0.4874} & \multirow{5}{*}{0.0305} &                                                                                                 & Shaanxi      & 0.3607           &                         &                         \\
                              & Chongqing & 0.5468           &                         &                         & \multirow{4}{*}{\begin{tabular}[c]{@{}l@{}}Middle reaches of \\ the Yangtze River\end{tabular}} & Anhui        & 0.4209           & \multirow{4}{*}{0.6014} & \multirow{4}{*}{0.053}  \\
                              & Sichuan   & 0.5221           &                         &                         &                                                                                                 & Jiangxi      & 0.9968           &                         &                         \\
                              & Guizhou   & 0.3076           &                         &                         &                                                                                                 & Hubei        & 0.4878           &                         &                         \\
                              & Yunnan    & 0.2939           &                         &                         &                                                                                                 & Hunan        & 0.5001           &                         &                         \\ \bottomrule
\end{tabular}
\end{table}
\end{landscape}

\subsubsection{Cluster Analysis Based on Variable Weighting Frontier}
\label{sec-05-03-02}

This section clusters provinces according to input and output variable weighting frontiers as different provinces achieve optimal efficiency. This study uses the K-means clustering method, determining the optimal cluster number based on the elbow rule. Table \ref{tab-07} shows the clustering results. The clustering result displays a profile coefficient of 0.707, a DBI of 0.318, and a CH of 162.582, indicating a favorable clustering effect. Among them, 30 provinces are classified into three categories with proportions of 63.333$\%$, 26.667$\%$, and 10$\%$, respectively. The variability in input and output variables indicates significant differences across clustering categories at the p-value of $0.000^{***}$. This study named the three categories as technology-driven provinces (TDP), development-balanced provinces (DBP), and transition-potential provinces (TPP) based on the centroid characteristics of the weights of the input and output variables in each cluster, as described in Table \ref{tab-08}.

\begin{table}[h]
\centering
\caption{K-means clustering results}
\label{tab-07}
\begin{tabular}{llllll}
\hline
\multirow{4}{*}{\begin{tabular}[c]{@{}l@{}}Input and \\ output variables\end{tabular}} & \multicolumn{3}{l}{\multirow{2}{*}{\begin{tabular}[]{@{}l@{}}Clustering categories \\ (mean ± standard deviation)\end{tabular}}}                                                                                                        & \multirow{4}{*}{F} & \multirow{4}{*}{P} \\
                                       & \multicolumn{3}{l}{}                                                                                                                                                                                                                     &                    &                    \\ \cline{2-4}
                                       & \multirow{2}{*}{\begin{tabular}[]{@{}l@{}}Category 1\\ (n=19)\end{tabular}} & \multirow{2}{*}{\begin{tabular}[]{@{}l@{}}Category 2\\ (n=8)\end{tabular}} & \multirow{2}{*}{\begin{tabular}[]{@{}l@{}}Category 3\\ (n=3)\end{tabular}} &                    &                    \\
                                       &                                                                              &                                                                             &                                                                             &                    &                    \\ \hline
$L$                                      & 0.105±0.024                                                                  & 0.121±0.006                                                                 & 0.047±0.005                                                                 & 15.621             & 0.000***           \\
$K$                                      & 0.117±0.017                                                                  & 0.173±0.009                                                                 & 0.05±0.008                                                                  & 79.475             & 0.000***           \\
$T$                                      & 0.512±0.017                                                                  & 0.183±0.043                                                                 & 0.44±0.004                                                                  & 442.12             & 0.000***           \\
$E$                                      & 0.063±0.032                                                                  & 0.109±0.005                                                                 & 0.046±0.004                                                                 & 10.126             & 0.001***           \\
$Y$                                      & 0.162±0.037                                                                  & 0.276±0.017                                                                 & 0.369±0.026                                                                 & 74.218             & 0.000***           \\
$C$                                      & 0.042±0.0                                                                    & 0.138±0.007                                                                 & 0.048±0.006                                                                 & 1845.7             & 0.000***           \\ \hline
\multicolumn{6}{l}{\small{Note: ***, **, * represent 1\%, 5\%, and 10\% significance levels, respectively.}}                                                                                                                                                                                                                       
\end{tabular}
\end{table}

\begin{table}[h]
\centering
\caption{Province classification}
\label{tab-08}
\begin{tabular}{ll}
\hline
Clustering category                             & Province                                                                                                                                                                                                      \\ \hline
Technology-driven province (TDP)       & \begin{tabular}[l]{@{}l@{}}Tianjin, Hebei, Neimenggu, Liaoning,\\ Heilongjiang, Jiangsu, Zhejiang, Anhui,\\ Shandong, Henan, Hubei, Hunan, Gansu\\ Chongqing, Sichuan, Guizhou, Yunnan\\ Shaanxi, Ningxia\end{tabular} \\
Development-balanced province (BDP)     & \begin{tabular}[l]{@{}l@{}}Beijing, Jilin, Shanghai, Fujian,\\ Jiangxi, Guangdong, Hainan, Qinghai\end{tabular}                                                                                                     \\
Transition-potential province (TPP) & Shanxi, Guangxi, Xinjiang                                                                                                                                                                                       \\ \hline
\end{tabular}
\end{table}

TDP covers Tianjin, Hebei, Neimenggu, and 14 other provinces. From the weights of each input and output variable, technical inputs are the most critical factor in evaluating ICEE among these provinces, with a weight of 0.512. Capital inputs, labor inputs, and industrial output follow it. For provinces in this category, technological innovation is essential to boosting industrial productivity and reducing carbon emissions. At the same time, the focus on industrial output in these regions demonstrates their necessity to balance the need to safeguard a certain economic output level in the emission reduction process. Thus, we define this category as a technology-driven province. Notably, among these regions, Neimenggu performs the best in ICEE and can be regarded as a benchmark of technology-driven provinces.

DBP includes eight provinces, including Beijing, Shanghai, and Jiangxi. These provinces are relatively close to each input and output ratio, with fluctuations ranging from 0.1 to 0.28. This indicates that the DBP region shows a balanced development in capital, labor, technology, and industrial output and is therefore classified as the development-balanced province. These regions emphasize the rational use of integrated resources, including labor, capital, technology, and energy, by optimizing the allocation of resources to achieve efficient operations and actively reducing carbon emissions. The ICEE values are generally high among the development-balanced provinces. Beijing, Shanghai, Jiangxi, Guangdong, and Hainan all have efficiency values over 0.99, which can be regarded as the benchmark provinces of the development-balanced provinces.

TPP includes the three provinces of Shanxi, Guangxi, and Xinjiang. These provinces have disadvantages in labor, capital, and energy consumption and lack of attention to carbon emission outputs, but are prominent in technological inputs and industrial outputs. Thus, it is clear that these provinces are now focusing on upgrading their technological inputs to promote technological innovation and improve the output quality. However, there is still a need to focus on using resources efficiently and protecting the environment in economic development, significantly reducing carbon dioxide emissions. Given this, this paper considers these provinces as transition potential provinces. Among them, the ICEE of Xinjiang is relatively excellent and can be regarded as an exemplary province of transition potential.

\section{Discussion}
\label{sec-06}

This section examines ICEE and clustering results based on weights of input and output variables across various policy preference scenarios. It also presents specific recommendations for carbon emission reduction policies and strategic measures tailored to different provinces. Methodologically, this analysis serves as a sensitivity assessment of ICEE under varying policy preferences. We permute the three policies given in Section \ref{sec-05-02} to generate six policy preference scenarios, outlined in Table \ref{tab-08}. Initially, we compute the ICEE under these scenarios through the same process shown in Section \ref{sec-05}. Subsequently, we conduct cluster analysis on the weight characteristics of input and output variables to identify provinces exhibiting similar efficiency frontier features when achieving the optimal ICEE in different contexts.

\begin{table}[h]
\caption{Policy preference settings}
\centering
\label{tab-09}
\begin{tabular}{ll}
\hline
Policy preference & Policy ranking \\ \hline
S1                          & P1 > P2 > P3 \\
S2                          & P1 > P3 > P2 \\
S3                          & P2 > P1 > P3 \\
S4                          & P2 > P3 > P1 \\
S5                          & P3 > P1 > P2 \\
S6                          & P3 > P2 > P1 \\ \hline
\end{tabular}
\end{table}

\begin{figure}[h]
    \centering
    \includegraphics[width=0.65\linewidth]{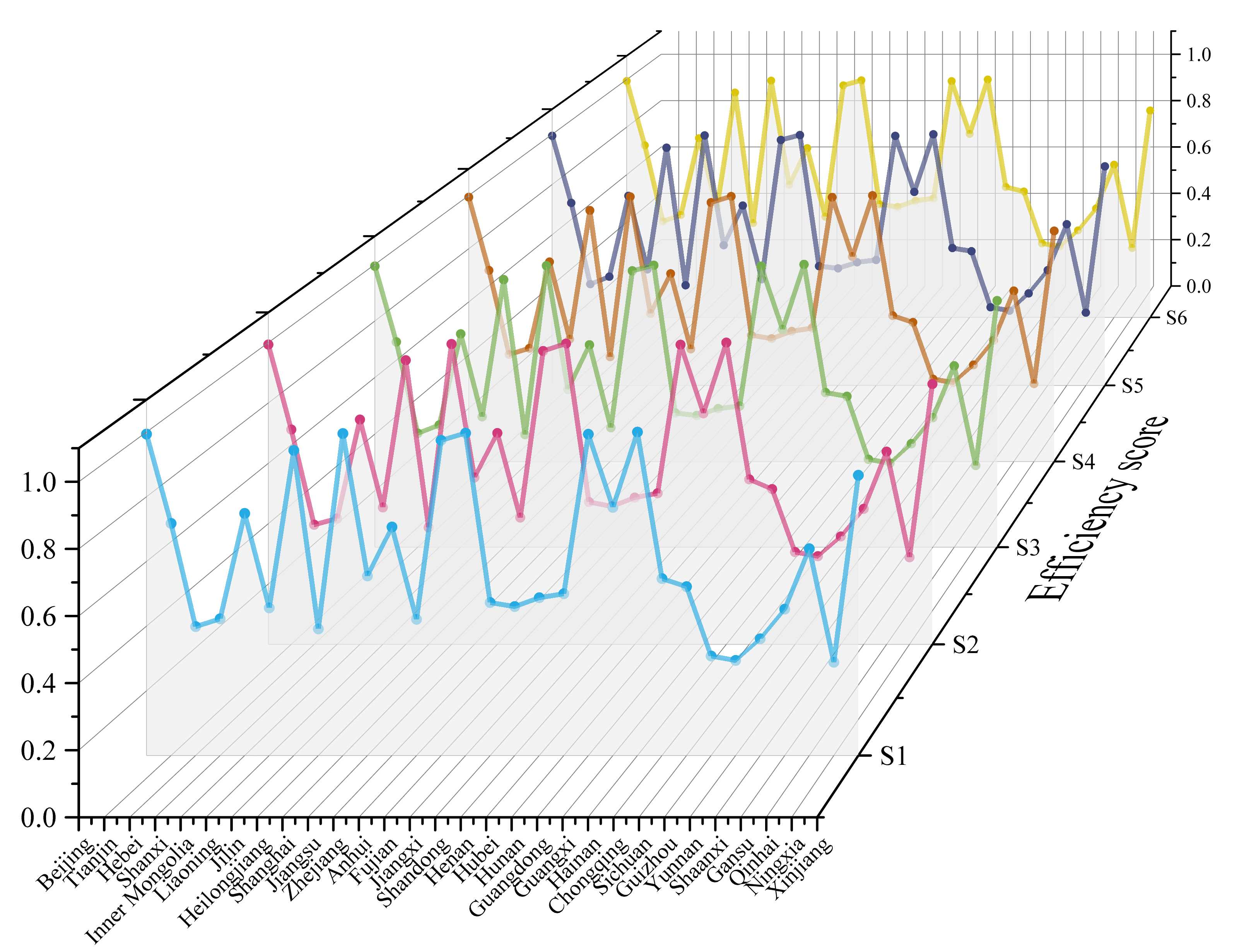}
    \caption{Technical efficiency of provinces under different policy preference scenarios}
    \label{fig-03}
\end{figure}

\begin{figure}[h]
    \centering
    \includegraphics[width=0.75\linewidth]{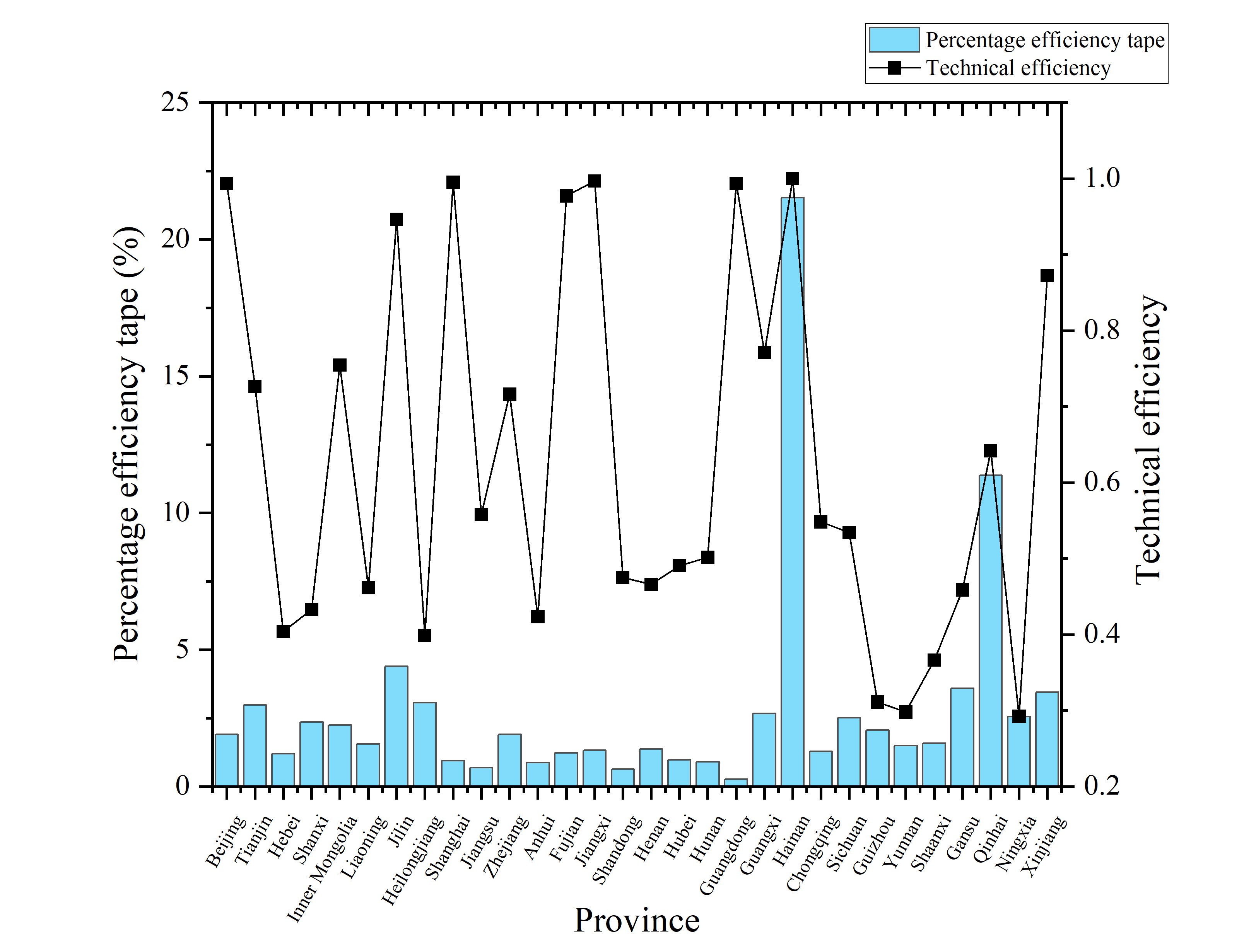}
    \caption{Optimal technical efficiency and percentage tape across different policy preference scenarios}
    \label{fig-04}
\end{figure}

Figure \ref{fig-03} illustrates the ICEE under different policy preference scenarios. The results show that the ICEE across provinces under different policy preference scenarios are generally similar. The national average efficiency is highest under S5 at 0.6267 and lowest under S2 at 0.6205. However, as seen from Figure \ref{fig-04}, there are significant differences in the sensitivities of ICEE across policy preference scenarios for the 30 provinces in response to input data uncertainty. Hainan shows the most significant discrepancy at 21.52$\%$, followed by Qinghai at 11.38$\%$, while Guangdong shows the most minor discrepancy at 0.26$\%$.

Table \ref{tab-08} shows the best policy preference options of provinces in each category and their corresponding efficiency performance. Notably, S1, S2, and S3 are not the best efficiency policy preference options for any province. In the technology-dominant policy preference scenario (i.e., S4 and S5), most provinces achieve the optimal level of ICEE, and S4 and S5 cover 23$\%$ and 67$\%$ of the total number of provinces, respectively. Under the environment-dominant policy preference scenario, Chongqing, Ningxia, and Hunan achieve optimal efficiency. Through K-means clustering analysis of indicator weights, regional groupings of provinces sharing similar efficiency frontiers is identified. Comparative analysis of optimal efficiency scores within each group pinpoint leading benchmark provinces in each category. The regional type of provinces is the same as in Section \ref{sec-05-03}. The TDP includes 18 provinces, such as Tianjin and Chongqing, whose average optimal efficiency score is 0.5432. Provinces above this average include Chongqing, Tianjin, Neimenggu, Jilin, Zhejiang, and Fujian. The TDP provinces have coverage ratios of $11$\%$, $28$\%$, and 61$\%$ under optimal policy scenarios S4, S5, and S6, respectively. Notably, Fujian, with an optimal efficiency score of 0.9773, stands out as the benchmark province in the TDP category. The BDP includes 9 provinces, such as Hunan, Beijing, and Shanghai, with an average optimal efficiency score of 0.7729. Provinces above this average include Beijing, Shanghai, Jiangxi, Guangdong, and Hainan. The BDP provinces have coverage ratios of 11$\%$, 78$\%$, and 11$\%$ under optimal policy scenarios S4, S5, and S6, respectively. Beijing, Shanghai, Guangdong, and Jiangxi provinces achieve the optimal ICEE values close to 1 and are recognized as benchmarks. The TPP includes three provinces: Shanxi, Xinjiang, and Guangxi, whose average optimal efficiency score is 0.6922. Except for Shanxi, the ICEE of all other provinces are higher than this average. The ratio of TPP provinces under policy preference scenarios S5 and S6 is 67$\%$ and 33$\%$, respectively. Xinjiang becomes the benchmark province in this category, with an efficiency score of 0.8723.

\begin{table}[h]
\centering
\caption{Optimal policy scenarios and corresponding efficiency for provinces under each category}
\label{tab-10}
\begin{tabular}{@{}llll@{}}
\toprule
\multicolumn{1}{c}{} &
  \multicolumn{1}{c}{S4} &
  \multicolumn{1}{c}{S5} &
  \multicolumn{1}{c}{S6} \\ \midrule
TDP &
  \begin{tabular}[c]{@{}l@{}}Chongqing, \\ Ningxia\end{tabular} &
  \begin{tabular}[c]{@{}l@{}}Tianjin, Neimenggu, Jilin, \\ Liaoning, Heilongjiang, Jiangsu,\\ Zhejiang, Fujian, Sichuan, \\ Guizhou, Gansu\end{tabular} &
  \begin{tabular}[c]{@{}l@{}}Hebei, Anhui, \\ Shandong, Hubei,\\ Yunnan\end{tabular} \\
BDP &
  Hunan &
  \begin{tabular}[c]{@{}l@{}}Beijing, Shanghai, Jiangxi, \\ Henan, Guangdong, Hainan, \\ Shaanxi\end{tabular} &
  Qinghai \\
TPP &
  $--$ &
  Shanxi, Xinjiang &
  Guangxi \\ \bottomrule
\end{tabular}
\end{table}

Based on the above results and the current economic status of each province, this study proposes policy recommendations for advancing `dual-carbon' strategy in three categories of provinces:
\begin{itemize}
    \item TDP: Provinces in TDP should actively promote research and development of low-carbon technologies, using government funding to adopt efficient production technologies and innovative processes to reduce energy consumption and emissions. Enterprises should invest in energy-saving equipment and intelligent control systems to improve energy use efficiency. At the same time, it should increase investment in clean energy sources such as solar and wind and reduce its dependence on fossil fuels. It should also introduce foreign energy-saving technologies and techniques to improve industrial energy efficiency through technological innovation. Technology-driven provinces such as Ningxia and Chongqing should focus on advanced technologies and pay special attention to the sustainable use and conservation of water, land, energy, and natural resources. Ningxia's industrial sector has low ICEE and needs to strengthen clean energy development and technological innovation. Ningxia and Qinghai are geographically similar, and they can establish a partnership to address environmental challenges. Coastal areas such as Jiangsu, Zhejiang, and Fujian are suggested to utilize offshore wave and wind energy to accelerate the construction of a clean, low-carbon, safe, and efficient multi-energy supply system. In high-carbon regions such as Neimenggu, Gansu, Jilin, Heilongjiang, and Guizhou, local governments should promote the innovation and application of new technologies, guide enterprises to focus on the development of new and emerging technology industries and adjust the energy structure to promote the substitution of fossil energy with cleaner, renewable and non-carbon energy sources. Provinces like Hebei, Anhui, Shandong, Hubei, and Yunnan traditionally depend on abundant local energy resources. In the future, these regions must optimize their industrial structure, enhance energy efficiency, unlock emission reduction potential, and foster green regional economic growth and energy-saving practices.
    \item BDP: Provinces in BDP aiming for balanced development must adopt comprehensive strategies to reduce industrial CO2 emissions. Adjusting policies for energy-intensive industries will optimize resource allocation and foster regional economic complementarity. Second, upgrading industries and maximizing resource utilization will enhance  CEE) ,exemplified by national park construction and pilot projects. Key provinces like Beijing, Shanghai, and Guangdong should leverage technological innovation to boost their economies and achieve harmonious economic and environmental coexistence. These regions should collaborate on technology and utilize innovative resources to establish an innovation-driven economic system, setting benchmarks for development. Beijing and Guangdong can lead in implementing comprehensive emission controls, while other areas should promote energy-saving technologies to transition to low-carbon industries and energy sources. Concurrently, enhancing partnerships with Guangxi and other regions in clean services will support upgrading low-carbon technologies and optimizing industrial structures across central, western, and northern regions. The technological and industrial revolution enables the central region's transition to green and sustainable development. Henan and Shaanxi should enhance investment in industrial technology, expand clean energy supply, and effectively manage energy and water resources. Qinghai must consider the environmental impact of economic development and prevent environmental damage during resource exploitation. Hunan should speed up industrial upgrading, shifting towards technology- and capital-intensive industries while fostering the growth of low-carbon and green sectors.
    \item TPP: Provinces in TPP, including Shanxi, Xinjiang, and Guangxi, differ economically from the eastern region and must prioritize future economic growth. They should actively promote the "dual-carbon" strategy to foster a low-carbon economic model. Addressing carbon-intensive industries through technological innovation and industrial upgrading is crucial for sustainable development. Xinjiang should utilize its abundant wind and solar resources to develop renewable energy industries as a production base. Shanxi and Guangxi should transition from traditional energy to green energy chemicals through innovation, industrial optimization, and enhanced energy efficiency. Developing the green coal chemical industry, advancing clean and efficient coal power technology, and eco-friendly coal mining are vital for achieving economic benefits and reducing carbon emissions. Coordination with regional development plans, considering differences in resource endowment and energy infrastructure, is essential for establishing a regional low-carbon spatial synergy development pattern.
\end{itemize}

\section{Conclusion}
\label{sec-07}

Given the severe challenges posed by global climate change, optimizing ICEE is central to achieving a low-carbon economic transformation. Significant differences in industrial carbon emissions across regions are influenced by factors such as regional industrial structure and resource endowments. Governments should, therefore, develop a framework for assessing ICEE, with specific policy preferences tailored to local circumstances. However, there is a gap in current research regarding evaluating CEE under policy variability and data uncertainty. Therefore, this study proposes $\delta$-SBM-OPA to address these challenges. Specifically, $\delta$-SBM constructs efficiency frontiers and their corresponding taps, deriving sensitivity indicators to address data uncertainty. Meanwhile, OPA provides a lower bound reference on the importance of input and output variables for the dual problem of $\delta$-SBM based on policy preference. The input and output variable weight frontiers of each DMU under various policy preference scenarios are analyzed using K-means clustering to categorize DMUs with similar efficiency frontiers. Additionally, the advanced benchmark DMU in each category is determined based on their optimal efficiency scores. The ICEE analysis of 30 provinces in China serves as an illustrative application of the proposed $\delta$-SBM-OPA model. Regarding policy preference settings, this study examines the influence of economic, environmental, and technological priority policies and corresponding policy preference on the ICEE. Furthermore, this study proposes the policy recommendation for carbon emission reduction. The key findings show that:
\begin{itemize}
    \item Regarding the policy preference of `economy > environment > technology', the average ICEE across 30 provinces in China is 0.6227, with only 12 provinces exceeding this value. The top five provinces, Beijing, Shanghai, Guangdong, Jiangxi, and Hainan, all have ICEE averages surpassing 0.99. Among the eight economic regions, the South Coast exhibits the highest mean ICEE at 0.9892 and the lowest variance at 0.0001. The ICEE values for the East Coast, North Coast, Middle Yangtze River, Northeast, and Northwest range from 0.7523 to 0.5617. Conversely, the Southwest and Northwest show the lowest ICEE values, with 0.4981 and 0.4874, respectively. Except for the South Coast, the ICEE variances in the other regions range from 0.0220 to 0.0607, indicating certain variability.
    \item The results of the ICEE under different policy preferences show that 27 provinces have the optimal ICEE in the technology-dominant policy preference scenario. Chongqing, Ningxia, and Hunan show the optimal ICEE under the environment-dominant policy preference scenario. The provinces can be categorized into technology-driven, development-balanced, and transition-potential based on their ICEE characteristics. Technology-driven provinces have 18 provinces with an average optimal ICEE of 0.5432, taking Fujian as the ICEE benchmark (0.9773). Of these, 89$\%$ are optimal under a technology-dominant policy preference and 11$\%$ under the environment-dominant policy preference. In the development-balanced provinces, 9 have an average optimal ICEE of 0.7729, with the ICEE benchmark nearing 1, including Beijing, Shanghai, Guangdong, and Jiangxi. This category is identical to the distribution of optimal policy preferences for technology-driven provinces, with 89$\%$ of technology-dominant and 11$\%$ of the environment-dominant. The transition-potential provinces, with an average optimal ICEE of 0.6922, include three provinces where Xinjiang serves as a benchmark with 0.8723, and all provinces achieve optimal under the technology-dominant policy preference. 
\end{itemize}

It is essential to highlight that the objective of this study is to develop a tool for analyzing CEE under varying policy preferences and data uncertainty. Therefore, this study employs single-year industrial carbon emission data from 30 provinces in China to illustrate the proposed model rather than using multi-year panel data. Correspondingly, in the future, multi-year panel data can be analyzed to investigate trends in CEE and variations in optimal policies across provinces. In addition, the proposed model can be extended to other industries, such as agriculture and tourism, to verify its rationality and applicability in assessing CEE. Finally, exploring the influence of policy preferences on CEE across different scales, such as cities and enterprises, represents a promising direction for future research.

\section*{Author Contributions}

\textbf{Shutian Cui}: Writing - original draft, Investigation, Conceptualization, Data curation, Visualization, Formal analysis. 
\textbf{Renlong Wang}: Supervision, Writing - original draft, Methodology, Conceptualization, Validation, Software. 

\section*{Declaration of competing interest}
The authors declare that they have no known competing financial interests or personal relationships that could have appeared to influence the work reported in this paper.

\section*{Data Availability}
Data will be made available on request.

\newpage

\bibliographystyle{elsarticle-harv} 
\bibliography{reference}





\end{document}